# Nuclear Transparency in the lightest nuclear interactions

M. Ajaz[1, 2], M. K. Suleymanov[1, 3], K. H. Khan[1], A. Zaman[1]
[1]Department of Physics, COMSATS Institute of Information Technology, Islamabad
[2]Department of Physics, Abdul Wali Khan University Mardan
[3]Joint Institute for Nuclear Research Dubna

**Abstract**

Some experimental results on nuclear transparency effect in pC- and dC-interaction at 4.2 A GeV/c (JINR Dubna) are presented. The "half angle" ($\theta_{½}$) technique was used and the particles with emission angle greater and less than $\theta_{½}$ are considered separately. The results of the experimental study have been compared with the simulation data coming from the Dubna Cascade model. The values of average multiplicity, average momentum, and average transverse momentum of charged pions and protons are analyzed as a function of the number of identified protons in an event. We observed some behaviors for the data which could be considered as some nuclear transparency effects. The lasts have been divided into three main groups depending on their probable behavior: leading effect; cascade effect; medium effect.

**Introduction:-**

Nuclear Transparency (NT) effect considered as an important phenomenon is connected with dynamics of hadron-nuclear and nuclear–nuclear interactions which could reflect some particular properties of the medium. Different mechanisms might lead to the appearance of the NT. Usually in an experiment the NT is defined as:

1. the ratio of nuclear cross section (σ) per nucleon to that on a free nucleon [1]

$$T_r = \frac{\frac{d\sigma}{dt}(pp \text{ quasi-elastic in nucleus})}{Z \frac{d\sigma}{dt}(pp \text{ elastic in hydrogen})}.$$

dσ/dt is the differential cross section of the process and Z is the charge of the nucleus.

2. the ratio of cross section for scattering measured in nuclear target to the cross section for scattering in (model) plane wave impulse approximation (PWIA) [2]. Numerically the ratio was given using the following expression.

$$T(Q^2) = \frac{\int_V d^3 p_m dE_m Y_{exp}(E_m, \vec{p}_m)}{\int_V d^3 p_m dE_m Y_{PWIA}(E_m, \vec{p}_m)},$$

Where the integral is over the phase space $V$ defined by the cuts $E_m < 80$ MeV, threshold energy and $|\mathbf{p}_m| < 300$ MeV/$c$, threshold 3-momentum. Whereas $Y_{exp}(E_m, \mathbf{p}_m)$ and $Y_{PWIA}(E_m, \mathbf{p}_m)$ are the corresponding experimental and simulation yields.

3. The ratio of experimental charged normalized yield (Y bar) divided by the charge normalized Mote Carlo equivalent yield ($Y_{SIMC}$ bar) for a target with nucleon number A to the same division that for hydrogen (H) target. [3]

$$T = \frac{(\bar{Y}/\bar{Y}_{SIMC})_A}{(\bar{Y}/\bar{Y}_{SIMC})_H},$$

4. Some behavior of the average values of secondary particles produced in hadron-nucleus (hA) and nucleus-nucleus (AA) collisions as a function of the number of g-particles [4, 5] or identified protons ($N_p$) [6] using the "half angle" ($\theta_{½}$) technique.

5. Some behavior of the nuclear modification factor ($R_{AA}$) as a functions of a number of participants nucleons ($N_{part}$) [7] $R_{AA}$ is considered as a best parameter to study the Color Transparency (CT), which was introduced for the first time by Brodsky and Mueller [8, 9] in 1982. CT is the prediction that hadrons produced in exclusive reactions with high four momentum transfers squared ($Q^2$) can pass through nuclear matter with reduced interactions [2, 10-12]. According to Quantum Chromo Dynamics (QCD) [13], hard exclusive processes select special configurations of the hadron wave function where all quarks are close together, forming a color neutral small size configuration (SSC) with transverse size $r_\perp \sim 1/Q$. The external color field in these SSCs vanishes because their color fields cancel each other as the distance between quarks diminishes. Quark Gluon Plasma (QGP) [14] is one of the main concern of the todays heavy ion colliders. The CT effect is considered as a signal on Quark Gluon Plasma (QGP) formation [15-17]

It is clear that all the aforementioned definitions of NT are very close to each other.

The idea of nuclear transparency at low energy based on the parameterization of Bethe [18] is given by the following expression [19].

$$\sigma_R = (r_o A^{1/3} + \lambda)^2 [1 - Zze^2/(R+\lambda)E_o][1-T]$$

Where $r_o$ is the effective reduced nuclear radius, $\lambda$ is the reduced wavelength of the incident particle, z and Z the charge of the incident particle and the target nucleus respectively and T is the transparency. The expression given above is valid for protons in the energy range from 200 to 800 MeV.

NT in Quasi-elastic A(p, 2p) reaction" was performed at Brookhaven National Laboratory (BNL) [1, 20, 21]. Results obtained at BNL are inconsistent with CT and can be explained in terms of nuclear filtering or charm resonance states [22-25]. The onset of Nuclear Transparency is Quasi-elastic A(e,e'p) reaction were carried out at SLAC [26, 27] and Thomas Jefferson National Accelerator Facility (Jlab) [28] the A-dependence of the above mentioned reaction has been studied with deuteron, carbon, iron, and Gold as target nuclei at momentum transfers ranging from $Q^2 = 1$ to 6.8 $(GeV/c)^2$. They found no substantial rise in the nuclear transparency within errors in any of the nuclei studied. D. Abbott et al.,[29], studied the A(*e*, *e'p*) reaction using Carbon, Iron and Gold as targets at momentum transferred squared varied from 0.6 to 3.3 $GeV^2$. They observed that the results on carbon as a target do not reveal any significant increase in the nuclear reduction in the energy range where the nucleon-nucleon total cross sections increase significantly as pion production begins to dominate.

All experiments, as described above (SLAC [26, 27], JLab [28]) were unsuccessful to produce evidence of CT even for high values of $Q^2$.

The disintegration of polarized and un polarized $^2H$ targets by electron d(e, e′p)n at high momentum transfer were studied by L. L. Frankfurt et al., [30], but no conclusive model independent evidence for CT has been observed for qqq system --baryons. An earlier onset of CT for mesons production than that for hadrons was suggested [31], as it is most probable to produce a small transverse size in a two quark (qq´) system than in a three quark (qqq) system. D. Dutta et al.,[32] measured the nuclear transparency of 4He(γn→pπ-) and concluded that the nuclear transparency results from this study deviated from the traditional nuclear physics picture. Furthermore the nuclear transparency effect measured form the study as a function of momentum transfer is in good agreement with Glauber calculations which include the QCD phenomenon of CT. B. Clasie et. al., [3] measured the cross section and hence the NT of the pion electro production from hydrogen, deuteron, carbon, copper and gold targets and compared his results with Glauber and Glauber+CT calculations [33] and Glauber+SRC+CT [34]. They found that the $Q^2$ and atomic number (A) dependence of the nuclear transparency show deviations from traditional Glauber calculations, and are consistent with calculations of CT. M. R. Adams et al. [35] measured the NT in exclusive incoherent ρ production from different nuclear targets and observed an increase in the NT. K. Ackerstaff et al. Hermes Collaboration [36] measured the exclusive incoherent electro production cross section of the ρ0 (770) meson and hence the transparency from Hydrogen, deuteron, Helium ($^3$He), and Nitrogen targets as a function of coherent length ($l_c$) of the interaction of qq´ fluctuation with the nuclear medium. they compared their study with some of the previous experiments, such as with [37] where they measured the transparency to incoherent ρ0 production with 4 and 8 GeV photons and the E665 collaboration at FNAL measured with 470 GeV muons [35] beams and with Glauber calculation of Hufner et al. for $^3$He and $^{14}$N [38]. The nuclear transparency was found to decrease with increasing coherence length of quark antiquark fluctuations of the virtual photon. The transparencies extracted from the data agree well with the previous measurements and models including high energy ISI and FSI. Hermes Collaboration [39] also studied the exclusive coherent and incoherent electro production of the ρ0 meson from hydrogen ($^1$H) and nitrogen ($^{14}$N) targets as a function of $l_c$ and $Q^2$. The NT was found to increase (decrease) with increasing coherence length for coherent (incoherent) ρ0 electro production. They observed a rise of NT with $Q^2$ for fixed coherence length, which is in agreement with theoretical calculations of CT. The CLASS collaboration [40] in search for the medium modification through the properties including mass and width of ρ meson produced the light vector mesons (ρ, ω, and φ) in $^2$H, $^3$H, $^{12}$C and $^{56}$Fe at normal nuclear densities and zero temperature. The results obtained by CLASS collaboration were found to be different from the KEK proton synchrotron measurement [41], where they detected ρ, ω, and φ mesons from the same decay channel but was produced with 12 GeV energy proton beam.

CT is considered as an important effect to get the information on particular properties of the Quark Gluon Plasma (QGP)[15, 16]. Different mechanisms for the apparent transparency of the sQGP at LHC are discussed in [42]. Theoretical possibilities are discussed that could contribute to the apparent transparency (decreased opacity) of the sQGP relative to the WHDG/DGLV (radiative+elastic+geometric fluctuation) jet energy loss model extrapolation from RHIC to LHC include: Baryon anomaly [43-45]; Gluon feedback [46]; Gluon to quark jet conversion [47]; Gluon self-energy [48, 49]; Is the jet-medium coupling reduced at LHC: $\alpha_s$(LHC) < $\alpha_s$(RHIC) ?

At high energies the study of nuclear transparency effect in hadron-nucleus and nucleus-nucleus collision for the first time was carried out using "half angle" ($\theta_{\frac{1}{2}}$) technique by [4, 5] . The value of the $\theta_{\frac{1}{2}}$ was defined as an angle which divides the particle multiplicity into two equal parts in nucleon-nucleon (NN) interaction. They studied the behavior of s-particles ( the particles with β>0.7 in the emulsion experiments (in the papers [4, 5] ) and a number of pions ( in the paper [6])  as a function of g-particles (the particles with 0.23≤ β<0.7 (in the paper [4, 5]) and a number of identified protons (Np) (in the paper [6]). They observed that with increasing the number of g-particles (or Np) the values of the average multiplicity of the inner cone s-particles [4, 5] or multiplicity of identified pions [6] did not change being approximately equal to the multiplicity of these particles for the pp-collisions. So it was claimed to be the observed "transparency". Thought the values of the average multiplicity of the out cone s-particles (or pions) decreased linearly with a number of the g-partivcles (or $N_p$). The g (or $N_p$)-dependences for the values of the average pseudorapidity ($<\eta>$)[4, 5] of inner cone s-particles or the values of the average momentum ($<p>$) for inner cone pions [6] demonstrated that the values of the $<\eta>$ (or $<p>$) decreases linearly with g (or $N_p$). So the observed transparency in the case of multiplicity could not be confirmed as total transparency. It means different mechanisms could be the reason of the nuclear transparency effect. Y. Afek et, al., [50] has broadly divided the various models that have been suggested so far for high energy particle-nucleus collisions and for high energy nucleus-nucleus collisions into two categories. The first category include all models which assume that particle-nucleus collisions is a multistep process which consist of independent collisions which nucleons encountered when propagating through a target nucleus. This category include Intra-nuclear Cascade Models [51], Leading particle Cascade Models [52-55], Energy Flux Cascade Models [56], Multiperipheral Regge Type Models [57, 58] and various types of Statistical and Hydrodynamical Models [59]. The second category included all models that assume that particle-nucleus collisions in a single step process where a few nucleons (or partons) in the nucleus interact collectively with the incident particle [60-65]. For first category of models as per definition given above transparency could appear as a result of simultaneous action of different effect which will not carry any information about particular properties of matter. For the second category of the models the collective response of the nucleons provides the information about some specific property of the medium.

That is why the main goal of the paper is to look for the transparency effect of nuclear matter and to understand whether the effect connects with the first category of models or does with second ones.

To reach the goal we applied "half angle" technique [4-6]. We defined "half angle" ($\theta_{½}$) to be the angle which equally divides the multiplicity of secondary charged particles produced in NN-collisions. The values of the $\theta_{½}$ was determined as $\theta_{½} = 25°$. Beside $25°$ we used $5°$, $10°$, $15°$, and $20°$. "Half angle" divides the particles into the inner and outercone. So the particles with $\theta < \theta_{½}$ are named as inner cone particles and the one with $\theta > \theta_{½}$ are termed as outer cone particles. We defined the NT as an effect at which the characteristics of hadron-nucleus and nucleus-nucleus collisions do not depend on a number of identified protons ($N_p$), because the last connects with baryon density of matter. Finally the results are compared with the data coming from Dubna version of cascade model. For this study we used lightest nuclear interaction starting like pC- and dC-interaction at 4.2 A GeV/c because of the following reasons. Lightest nuclear systems are important link between nucleon-nucleon and nucleus-nucleus collisions. Comparing the results coming from the nucleon-nucleon, nucleus-nucleus and lightest nuclei, reactions at high energies are necessary to understand how nuclear transparency effect appears and how they depend on the characteristics of the medium.

As we have mentioned above different mechanisms could be reason of the NT. To separate the influences of the models which assume that particle-nucleus collisions is a multistep process and to get the information on transparency as some particular property of the medium we have compared the experimental data with one coming from the Dubna version of Cascade model[66-68].

**Experimental procedure and equipment**

The experimental data have been obtained from the 2-m propane bubble chamber of LHE, JINR. The chamber was placed in a 1.5 T magnetic field, and was exposed to beams of light relativistic nuclei at the Dubna Synchrophasotron. Practically all secondaries emitted at a $4\pi$ total solid angle were detected in the chamber. All negative particles, except identified electrons, were considered as $\pi^-$ − mesons. And this is justified as the contaminations by misidentified electrons and negative strange particles do not exceed 5% and 1%, respectively. The average minimum momentum for pion registration was set to about 70 MeV/c. The protons were selected by the statistical method applied to all positive particles with momentum of p ~150 MeV/c (we identified slow protons with p≤700 Mev/c by ionization in the chamber). In this experiment, we used 12757 pC, 9016 dC, interactions at a momentum of 4.2 A GeV/c ( for methodical details see[69]). In the case of cascade code we used 50000 pC-interactions at the same energy.

**Results**

A. **Average characteristics of inner cone protons in pC and dC interactions.**

The values of inner cone protons' average multiplicity $<n^{in}_p>_{pC}$, average momentum $<p^{in}_p>_{pC}$, and average transverse momentum $<p_T^{in}{}_p>_{pC}$ from experimental data in pC collision at 4.2A GeV/c (left hand side from top to bottom respectively) and from Cascade model (right hand side from top to bottom respectively) as a function of number of identified protons are shown in figure 1. The behavior is given for values of $\theta_{½} = 5^o, 10^o, 15^o, 20^o$ and $25^o$.

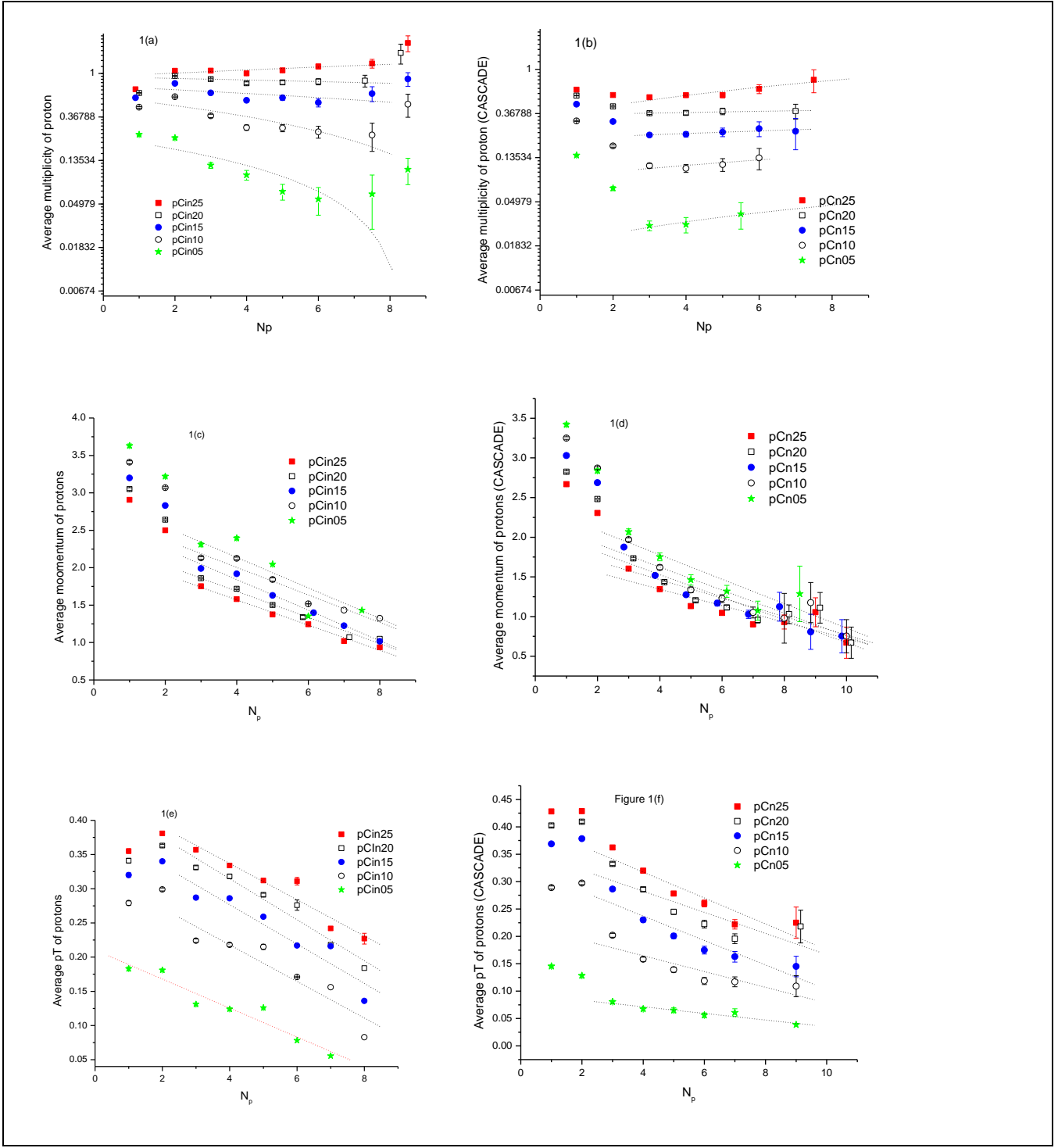

Figure 1 The average multiplicity, average momentum and average transverse momentum for the inner cone protons emitted in the pC-interactions at 4.2A GeV/c (left hand side from top to bottom) and simulated by the Cascade model (right hand side from top to bottom) as a function of number of identified protons with $\theta_{½}=25$(red square), $\theta_{½}=20$(open square), $\theta_{½}=15$(Blue triangles), $\theta_{½}=10$(open triangles) and $\theta_{½}=05$(Green stars).

Figure 1(a) demonstrates the $N_p$-dependence of the $<n^{in}_p>_{pC}$ at different values of the $\theta_{½}$ for experimental data. The behavior of $<n^{in}_p>_{pC}$ at half angle $\theta_{½}=25^0$ doesn't depend on $N_p$ in the region of $N_p$=2-9 having a very slight positive slope indicated from the fitting data using linear function (see table 1). The $N_p$-dependence demonstrates clearly a transparency for these protons. Same behavior can be observed for the values of the $\theta_{½}$ =15 and $\theta_{½}$ = $20^0$, but the values of $<n^{in}_p>_{pC}$ become less than 1. The values of $<n^{in}_p>_{pC}$ strongly depend on the $N_p$ at the values of $\theta_{½}$ =$10^0$ and $5^0$. So the transparency disappears below $15^o$. Fig1(b) demonstrated behavior for the values of the $<n^{in}_p>_{pC}$ at different $\theta_{½}$ coming from the Dubna Cascade code [66-68]. One can see that the values of the $<n^{in}_p>_{pC}$ are systematically and essentially less than 1 and less than the values for the data coming from the experiment. There are two regions for the $N_p$-dependences of the $<n^{in}_p>_{pC}$. In first region - $N_p$=1-3 the values of the $<n^{in}_p>_{pC}$ decrease with Np. In the second region the values of the $<n^{in}_p>_{pC}$ don't depend on the values of the $N_p$ and demonstrates transparency.

So we could say that the experimental data demonstrate clearly transparency for the protons with θ less than $25^0$: the $<n^{in}_p>_{pC} \cong 1$; $<n^{in}_p>_{Np}/<n^{in}_p>_2 \cong 1$ and don't depend on the values of $N_p$. The two lasts were observed for the protons with θ less than $20^0$ and $15^0$. The Cascade code simulation could not describe completely the experimental result: the $<n^{in}_p>_{pC} < 1$; the values of the $<n^{in}_p>_{pC}$ don't depend on the $N_p$ in the region of $N_p$> 3. (The results of the presented data in Fig.1(a) and 1(b) are approximated by linear function y=A+B*$N_p$, (here A and B are free parameters) conformed the conclusions, see Table 1).

The Fig.1(c) and 1(d) demonstrate the values of the $<p^{in}_p>_{pC}$ as a function of the $N_p$ for the experimental and code data respectively. There are two regions for the behaviors of the $<p^{in}_p>_{pC}$. In the first region ($N_p$=1-3) the values of $<p^{in}_p>_{pC}$ decreases sharply and in the second one the values of $<p^{in}_p>_{pC}$ decreases gradually with $N_p$. No transparency is observed in this case and the code data gives about the same behavior as the experimental one.

The values for the $<p_T^{in}_p>_{pC}$ as a function of the $N_p$ for the experimental and code data are given in Fig.1(e) & 1(f) respectively. One can see some oscillations for the behaviors of the experimental values of the $<p_T^{in}_p>_{pC}$ which get became weaker with decrease in $\theta_{½}$. Looking at the slope of the graphs (see Table 1) it is clear that the slope of the graphs increases with increasing "half angle". The slope is the least for $5^o$. So we could say that there are two regions for the behaviors of the $<p_T^{in}_p>_{pC}$ for the data coming from the code. In the first region ($N_p$=1-3) the values of $<p_T^{in}_p>_{pC}$ decrease sharply and in the second one the values of $<p_T^{in}_p>_{pC}$ decrease slowly with $N_p$.

Our claim of the observed transparency for the inner cone protons' average multiplicity could be explained in terms of leading effect. Leading particles are projectiles which could give some part of their energy during interaction [70]. The particles will have maximum energy in an event and would be identified in an experiment as inner cone particles due to their high energy /low angle. Having high energy they are able to pass by the medium very fast and save the large fraction of their initial energy. That is why medium seems transparent to them. This explanation is based on the fact that although the values of average multiplicity of the particles remain the same but their average momentum and average transverse momentum have been decreased.

Table1. The values of the parameter B (slope of the line) for inner cone protons in pC-interactions

| $\theta_{1/2}$ | <n> | | <p> | | <p_T> | |
|---|---|---|---|---|---|---|
| | Experiment | cascade | Experiment | cascade | Experiment | Cascade |
| 5 | -0.027±0.004 | 0.004±0.001 | -0.2±0.1 | 0.15±0.07 | -0.012±0.005 | 0.00603±0.001 |
| 10 | -0.053±0.008 | 0.008±0.004 | -0.18±0.0 | -0.14±0.03 | -0.027±0.006 | -0.01441±0.004 |
| 15 | -0.029±0.007 | 0.007±0.003 | -0.20±0.01 | -0.14±0.02 | -0.029±0.005 | -0.02239±0.004 |
| 20 | -0.016±0.005 | 0.005±0.002 | -0.17±0.01 | -0.12±0.02 | -0.03±0.003 | -0.01924±0.006 |
| 25 | 0.04±0.06 | 0.06±0.01 | -0.168±0.008 | -0.10±0.02 | -0.026±0.004 | -0.02366±0.005 |

Figure 2 given below shows the average values of inner cone protons' multiplicity $<n^{in}_p>_{dC}$, momentum $<p^{in}_p>_{dC}$, and transverse momentum $<p_T^{in}_p>_{dC}$ from the experimental data in dC - interactions at 4.2A GeV/c (left hand side from top to bottom respectively) and from the Cascade model (right hand side from top to bottom respectively) as a function of $N_p$. The behavior is studied for different values of the $\theta_{1/2}$ including $\theta_{1/2}$ = 5°, 10°, 15°, 20° and 25° as are given in the figures below. Figure 2(a) demonstrates the $N_p$-dependence of $<n^{in}_p>_{dC}$ at different values of the $\theta_{1/2}$ for experimental data. The behavior of $<n^{in}_p>_{dC}$ at $\theta_{1/2}=25^0$ is having a slight positive slope as was the case in pC-interactions (see Table 2 which demonstrates the result of fitting for parameter B the data presented in Figure 2 by linear function $A+B*N_p$). The $N_p$-dependence demonstrates some transparency for these protons in the region of $N_p \geq 3$ and having values greater than 1 as in comparison with 1(a) where the values were equal to 1. Same behavior can be observed for the values of the $\theta_{1/2}$ =15 and $\theta_{1/2}$ = 20° i.e. no dependence on $N_p$, but here the values of $<n^{in}_p>_{dC}$ for $\theta_{1/2}$ = 20° are still greater than 1 and for $\theta_{1/2}$ =15 the values are equal to 1. The values of $<n^{in}_p>_{dC}$ strongly depend on the $N_p$ at the values of $\theta_{1/2}$ =10° and 5° as was the case in pC data in figure 1. So the transparency disappears below 15°. Fig 2(b) demonstrates behavior for the values of the $<n^{in}_p>_{dC}$ at different half angles coming from the Dubna Cascade code [66-68]. One can see that the values of the $<n^{in}_p>$ are less than 1 as was observed in the previous case of figure 1(b) and less than the values for the data coming from the experiment. There are again two regions for the $N_p$-dependences of the $<n^{in}_p>_{dC}$. In first region - $N_p$=1-2 the values of the $<n^{in}_p>$ increases with Np. In the second region the values of the $<n^{in}_p>_{dC}$ don't depend on the values of the $N_p$ and demonstrates transparency.

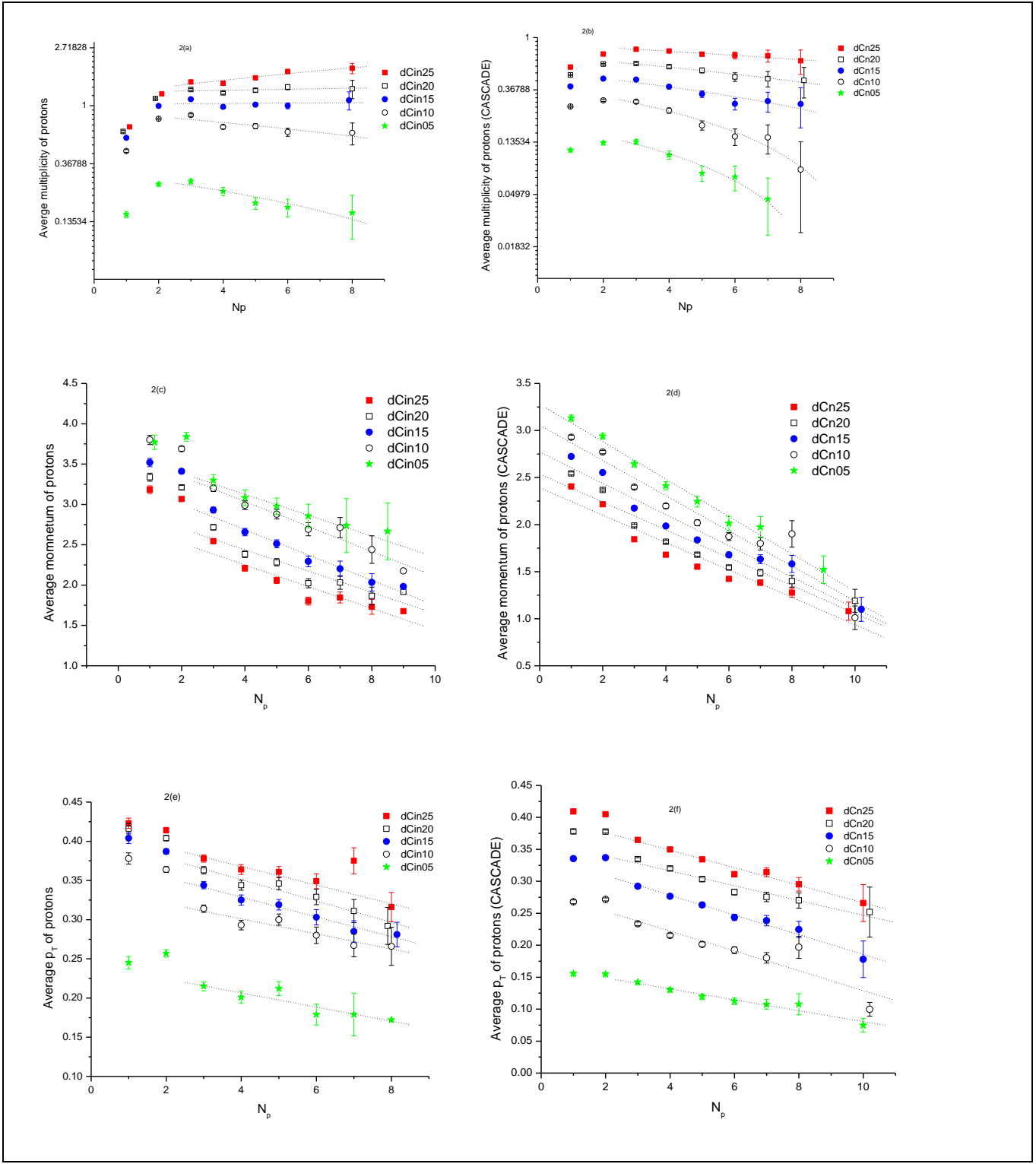

Figure 2 The average multiplicity, average momentum and average transverse momentum from experimental data in dC collision at 4.2A GeV/c (left hand side from top to bottom) and from the Cascade model in dC-interactions at 4.2A GeV/c (right hand side from top to bottom) as a function of number of identified protons with with $\theta_{½}$=25(red square), $\theta_{½}$=20(open square), $\theta_{½}$=15(Blue triangles), $\theta_{½}$=10(open triangles) and $\theta_{½}$=05(Green stars) as indicated in the figures.

So we could say that for inner cone protons emitted in dC-interactions the transparency observed too but compared with pC-interactions it start from the values of $N_p=3$. The last could be considered as some confirmation that the source of the transparency for inner cone protons could be the leading effect.

The Fig.2(c) & 2(d) demonstrate the values for $<p^{in}_p>_{dC}$ as a function of the $N_p$ for the experimental and code data. The values of $<p^{in}_p>_{dC}$ decreases in both cases with $N_p$.

Fig.2(e) & 2(f) demonstrate the values for the $<p_T^{in}_p>_{dC}$ as a function of the $N_p$ for the experimental and code data respectively. The value of average $p_T$ decreases with $N_p$ in the two cases as was the case in pC data of figure1.

Table 2. The values of the parameter B for inner cone protons emitted in dC-interactions at 4.2 AGeV/c

| $\theta_{½}$ | <n> | | <p> | | <$p_T$> | |
|---|---|---|---|---|---|---|
| | experiment | cascade | experiment | cascade | experiment | cascade |
| 5 | -0.022 ±0.005 | -0.022±0.003 | -0.13±0.01 | -0.20±0.01 | -0.009±0.002 | -0.009±0.001 |
| 10 | -0.039 ±0.01 | -0.04±0.003 | -0.15±0.01 | -0.19±0.02 | -0.010±0.002 | -0.016±0.003 |
| 15 | 0.00249 ±0.02 | -0.034±0.003 | -0.17±0.01 | -0.17±0.01 | -0.013±0.001 | -0.015±0.001 |
| 20 | 0.01±0.01 | -0.035±0.003 | -0.13±0.02 | -0.15±0.02 | -0.014±0.002 | -0.012±0.001 |
| 25 | 0.09±0.02 | -0.029±0.003 | -0.13±0.02 | -0.15±0.01 | -0.012±0.001 | -0.014±0.001 |

**B. Average characteristics of inner cone π⁻-mesons' in pC- and dC- interactions.**

The average values of the inner cone π⁻-mesons' multiplicity, momentum and average transverse momentum are given in figure 3(a-f) as a function of $N_p$ (the designations are same with figures 1 and 2). The behavior is studied at different values of the $\theta_{½}$ as given below. The values of $<n^{in}_{\pi^-}>_{pC}$ at different $\theta_{½}$ as a function of the $N_p$ are shown in Fig.3(a) for experimental data and 3(b) for the data coming from the Cascade code. The $<n^{in}_{\pi^-}>_{pC}$ as a function of the $N_p$ for both experimental and code data has about the same linear behavior. The slope of the graphs decreases with decreasing $\theta_{½}$ and seems to saturate at $5^o$ (showing some transparency) but there is still some positive slope in the case of model results. The degree of transparency increases with decreasing the value of $\theta_{½}$. The above qualitative description is justified by the fitting of the two graphs. The data were fitted by linear function $y = A+B*N_p$. The results for the values of parameter B demonstrate (see Table 3) that the model could describe satisfactorily the experimental data. This shows that the behavior in the two cases is the same within the errors.

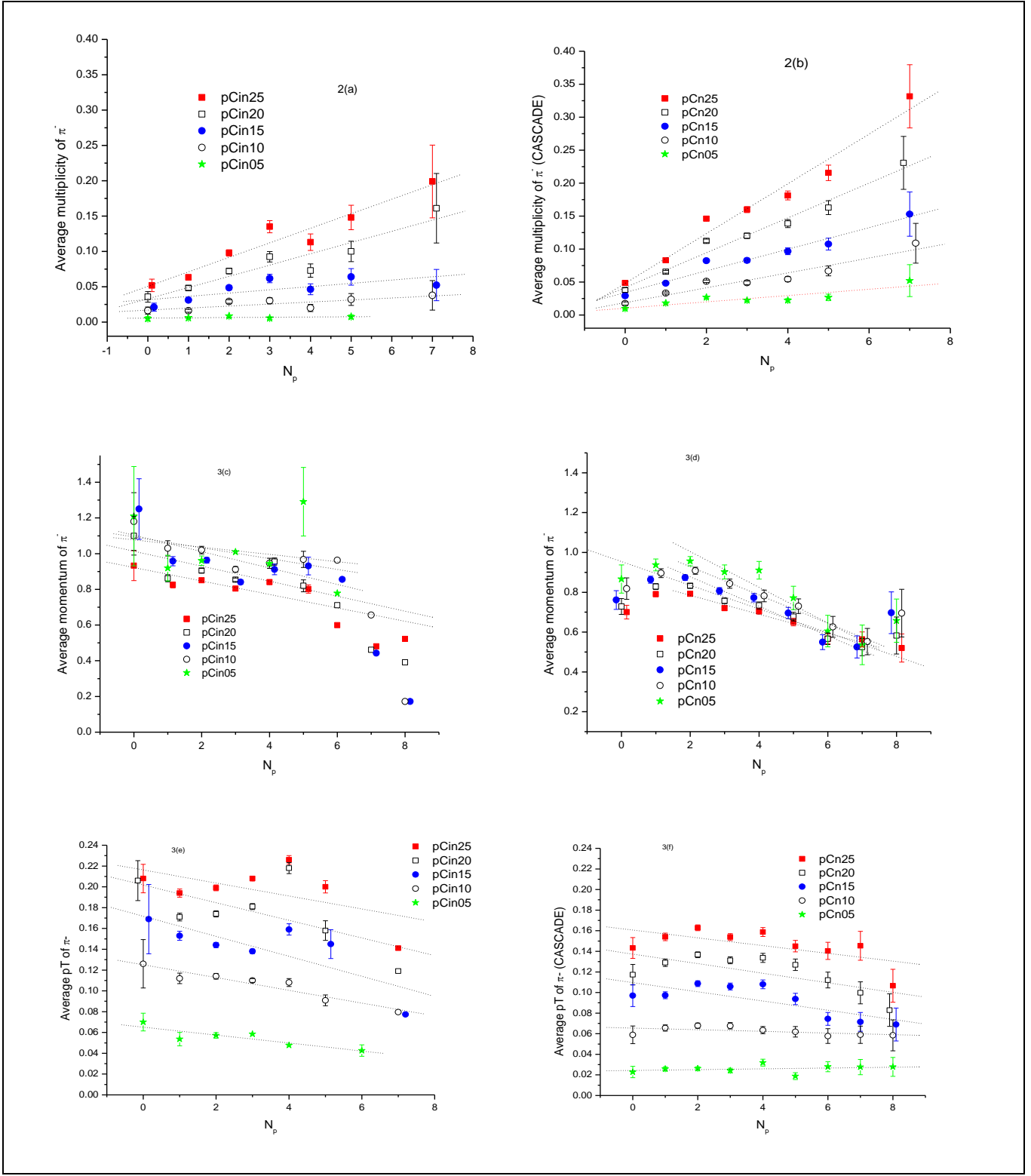

*Figure 3 the average multiplicity, average momentum and average transverse momentum of π⁻ - meson innercone from experimental data in pC collision at 4.2A GeV/c (left hand side from top to bottom) and from Cascade model in pC collision at 4.2A GeV/c (right hand side from top to bottom) as a function of number of identified protons with* with $\theta_{½}$=25(red square), $\theta_{½}$=20(open square), $\theta_{½}$=15(Blue triangles), $\theta_{½}$=10(open triangles) and $\theta_{½}$=05(Green stars) *as indicated in the figures.*

The values of $<p^{in}_{\pi^-}>_{pC}$ as a function of the $N_p$ for the experimental and code data are given in Fig3(c) & 3(d) respectively. There are two regions for the behavior of the $<p^{in}_{\pi^-}>_{pC}$. In the first region ($N_p$=0-4) the values of $<p^{in}_{\pi^-}>_{pC}$ don't depend on the $N_p$ (for experimental and model data, showing some transparency) and in the second one (for the values of $N_p > 4$) the values of $<p^{in}_{\pi^-}>_{pC}$ decrease with $N_p$. The data were fitted by linear function $y = A+B*N_p$. The results for the values of parameter B demonstrate that the model could describe satisfactorily the experimental data (see Table 3). The values for the $<p_T^{in}_{\pi^-}>_{pC}$ as a function of Np for the experimental and code data are shown in 3(e) & 3(f) respectively. The behavior of $<p_T^{in}_{\pi^-}>_{pC}$ as a function of $N_p$ is the same in the two cases. The experimental as well as Cascade data demonstrate some oscillations which increase with increasing $\theta_{\frac{1}{2}}$. We cannot determine the degree of oscillation but the values of $<p_T^{in}_{\pi^-}>_{pC}$ is higher in case of experimental data as compared to code data. Both the figure were fitted using linear function (see Table 3 for the results of parameters B) as given above which shows the similarity in the variation of the $<p_T^{in}_{\pi^-}>_{pC}$ as a function of $N_p$.

Now to summarize the discussion, we can conclude that the behavior of $<n^{in}_{\pi^-}>_{pC}$, $<p^{in}_{\pi^-}>_{pC}$ and $<p_T^{in}>_{pC}$ from experimental data (left panel of the figures) is well verified by the cascade model (right panel of the figures). As is discussed above Cascade model is one of the models which assume particle nucleus collision to be a multi-step process, which consists of successive independent collisions with nucleons encountered when propagating through a target nucleus. Although we have observed some transparency for lower values of $\theta_{\frac{1}{2}}$ but such transparency do not carry any information on the particular properties of the medium. This is due to the fact that cascade do not assume any medium properties. The effect is connected to some mechanism related to cascade model only.

Table 3. The values for the parameter B of inner cone $\pi^-$-mesons at different values of the $\theta_{\frac{1}{2}}$

| $\theta_{\frac{1}{2}}$ | <n> | | <p> | | <p_T> | |
|---|---|---|---|---|---|---|
| | Experiment | Cascade | Experiment | Cascade | Experiment | Cascade |
| 5 | 0.0004±0.0004 | 0.005±0.001 | -0.02±0.04 | -0.09±0.01 | 0.004±0.001 | 0.0004±0.0005 |
| 10 | 0.003 ±0.001 | 0.011±0.001 | -0.03±0.01 | -0.071±0.004 | -0.0061±0.0009 | -0.0008±0.0005 |
| 15 | 0.005 ±0.002 | 0.016±0.001 | -0.05±0.02 | -0.074±0.008 | -0.010±0.003 | -0.004±0.001 |
| 20 | 0.016±0.003 | 0.026±0.002 | -0.04±0.02 | -0.060±0.005 | -0.009±0.005 | -0.005±0.002 |
| 25 | 0.020±0.002 | 0.038±0.003 | -0.04±0.01 | -0.047±0.005 | -0.006±0.004 | -0.004±0.002 |

The values of the $\pi^-$- mesons' inner cone, average multiplicity $<n^{in}_{\pi^-}>_{dC}$, average momentum $<p^{in}_{\pi^-}>_{dC}$

and average transverse momentum $<p_T^{in}{}_{\pi-}>_{dC}$ from experimental data in dC collision at 4.2 A GeV/c (left hand side from top to bottom respectively) and from Cascade model (right hand side from top to bottom respectively) as a function of $N_p$ at different values of the $\theta_{\frac{1}{2}} = 5^o, 10^o, 15^o, 20^o$ and $25^o$ are given in the figures 4.

The values of the $<n^{in}{}_{\pi-}>_{dC}$ at different $\theta_{\frac{1}{2}}$ as a function of $N_p$ are shown in Fig. 4(a) for experimental data and 4(b) for Cascade data. Same is the case with dC as was the case with pC. The $<n^{in}{}_{\pi-}>_{dC}$ as a function of the $N_p$ for both experimental and code data has about the same linear behavior. The slope of the graphs decreases with decreasing $\theta_{\frac{1}{2}}$ and seems to saturate at $5^o$ (showing some transparency). The above observations are verified by the fitting of the two graphs.

The data were fitted by linear function $y = A+B*N_p$. The results fot the parameters B demonstrate that the model could describe satisfactorily the experimental data (see results in Table 4). This shows that the behavior in the two cases is the same within the errors. Comparing the behavior with figure 3(a) and 3(b) of pC data it is clear the behavior in the latter case is the same as the former one.

The values for $<p^{in}{}_{\pi-}>_{dC}$ as a function of the $N_p$ for the experimental and code data are given in Fig 4(c) and 4(d) respectively. The data were fitted by linear function $y = A+B*N_p$ and the results for the parameters B demonstrate (see Table 4) that the model could describe satisfactorily the experimental data. Comparison with pC-interactions' data in figure 3(c) and 3(d) one can see that the dC-interaction's data reinstate the previous results.

The Fig 4(e) & 4(f) demonstrate the values for $<p_T^{in}{}_{\pi-}>_{dC}$ in dC-collision as a function of $N_p$ for the experimental and code data respectively. One can see that the behavior of $<p_T^{in}{}_{\pi-}>_{dC}$ as a function of $N_p$ in case of experimental data has some negative slope while that of the Cascade code has no slope or very little negative slope. The experimental as well as the Cascade data demonstrate some oscillations which increase with increasing $\theta_{\frac{1}{2}}$. The degree of oscillation is not determined but the values of $<p_T^{in}{}_{\pi-}>_{dC}$ is higher in case of experimental data as compared to code data. Both the figure were fitted using linear function as before which shows the similarity in the variation of the $<p_T^{in}{}_{\pi-}>_{dC}$ as a function of $N_p$.

Comparing the results of pC data from figures 3 and dC data from figures 4 it is clear that the dC-interactions' data restore the results of pC-interactions' data, so we could say the observed transparency for the inner cone $\pi^-$ -mesons in dC-interactions at 4.2 A GeV/c could be a result of some cascade mechanisms of particle production.

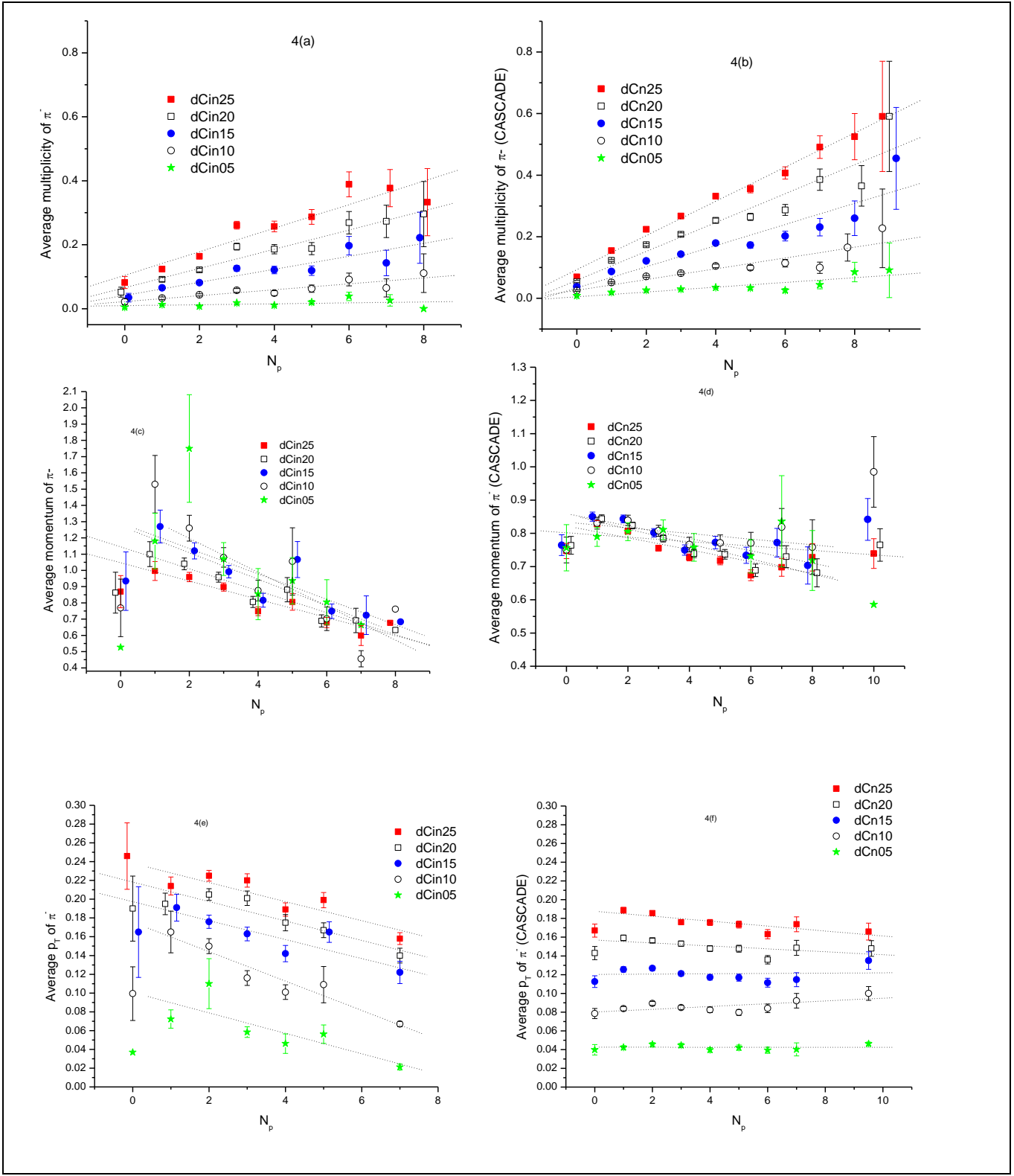

Figure 4 the average multiplicity, average momentum and average transverse momentum from experimental data in dC collision at 4.2A GeV/c (left hand side from top to bottom) and from Cascade model in dC collision at 4.2A GeV/c (right hand side from top to bottom) as a function of number of identified protons with $\theta_{1/2}$=25(red square), $\theta_{1/2}$=20(open square), $\theta_{1/2}$=15(Blue triangles), $\theta_{1/2}$=10(open triangles) and $\theta_{1/2}$=05(Green stars) as indicated in the figures.

Table 4 the values for the parameter B of inner cone $\pi^-$ -mesons in dC-interactions at 4.2 AGeV/c.

| $\theta_{\frac{1}{2}}$ | <n> | | <p> | | <p_T> | |
|---|---|---|---|---|---|---|
| | experiment | cascade | experiment | cascade | experiment | cascade |
| 5 | 0.001±0.001 | 0.008±0.002 | 0.0086±0.0002 | 0.008±0.002 | -0.011±0.004 | -0.00014±0.0003 |
| 10 | 0.009 ±0.002 | 0.012±0.003 | 0.0168±0.0006 | 0.015±0.002 | -0.016±0.002 | 0.0015±0.0007 |
| 15 | 0.020 ±0.003 | 0.034±0.005 | 0.027±0.004 | 0.018±0.002 | -0.010±0.003 | 0.0006±0.0009 |
| 20 | 0.030±0.003 | 0.047±0.006 | 0.037±0.006 | 0.029±0.003 | -0.010±0.002 | -0.0016±0.0008 |
| 25 | 0.037±0.005 | 0.056±0.002 | 0.049±0.008 | 0.044±0.006 | -0.010±0.002 | -0.0026±0.0007 |

### C. Average characteristics of outer cone $\pi^-$-mesons' in pC and dC interactions.

Figure 5 given below shows the average values of outer cone $\pi^-$ -mesons' multiplicity, momentum and transverse momentum as a function of $N_p$ at different values of $\theta_{\frac{1}{2}}$. The designations are same with previous figures.

The values of the $<n^{out}_{\pi^-}>_{pC}$ at different half angle $\theta_{\frac{1}{2}}$ as a function of the $N_p$ are shown in Fig.5(a) and 5(b) for experimental data and cascade respectively. The values of $<n^{out}_{\pi^-}>_{pC}$ increases linearly with $N_p$ in both the cases, but the average multiplicity in case of Cascade model has a steeper slope as compared to experimental data (see Table 5). Increasing $\theta_{\frac{1}{2}}$ the slope in the two cases decreases.

The values for the $<p^{out}_{\pi^-}>_{pC}$ as a function of the $N_p$ for the experimental and code data is demonstrated in Fig. 5(c) and 5(d) respectively. There is a big difference between the behavior of the experimental and the code data. Experimental data decrease linearly with $N_p$ and the slope of lines depend on $\theta_{\frac{1}{2}}$. At $\theta_{\frac{1}{2}} = 25^0$ we could see that the slope of line become minimum – transparency (see Table 5).

The increase in the value of transparency with increasing $\theta_{\frac{1}{2}}$ is clearly reflected by the results of experimental data. The code data demonstrate the existing of two regions for the behaviors of the $<p^{out}_{\pi^-}>_{pC}$. In the first region ($N_p$=0-5) the values of $<p^{out}_{\pi^-}>_{pC}$ decrease sharply and in the second one the values decrease slowly with $N_p$.

The Fig. 5(e) and 5(f) demonstrate $<p_T^{out}_{\pi^-}>_{pC}$ as a function of the $N_p$ for the experimental and code data respectively. One can see a drastic differences between experimental and code data. Experimental data shows some linear behavior with almost zero slopes whereas code data shows two regions. So we could say that the experimental data indicate some transparency. There is still some oscillation but the degree of oscillation is small and that is why the behavior could be considered as no dependence on $N_p$. transparency is observed for all values of $\theta_{\frac{1}{2}}$ in $<p_T^{out}_{\pi^-}>_{pC}$ in case of experimental data. The code data demonstrate two regions: in the first region the $<p_T^{out}_{\pi^-}>_{pC}$ value decreases sharply whereas in the second region the values decrease slowly. There is different behavior for different $\theta_{½}$ as a function of $N_p$ (see Table 5).

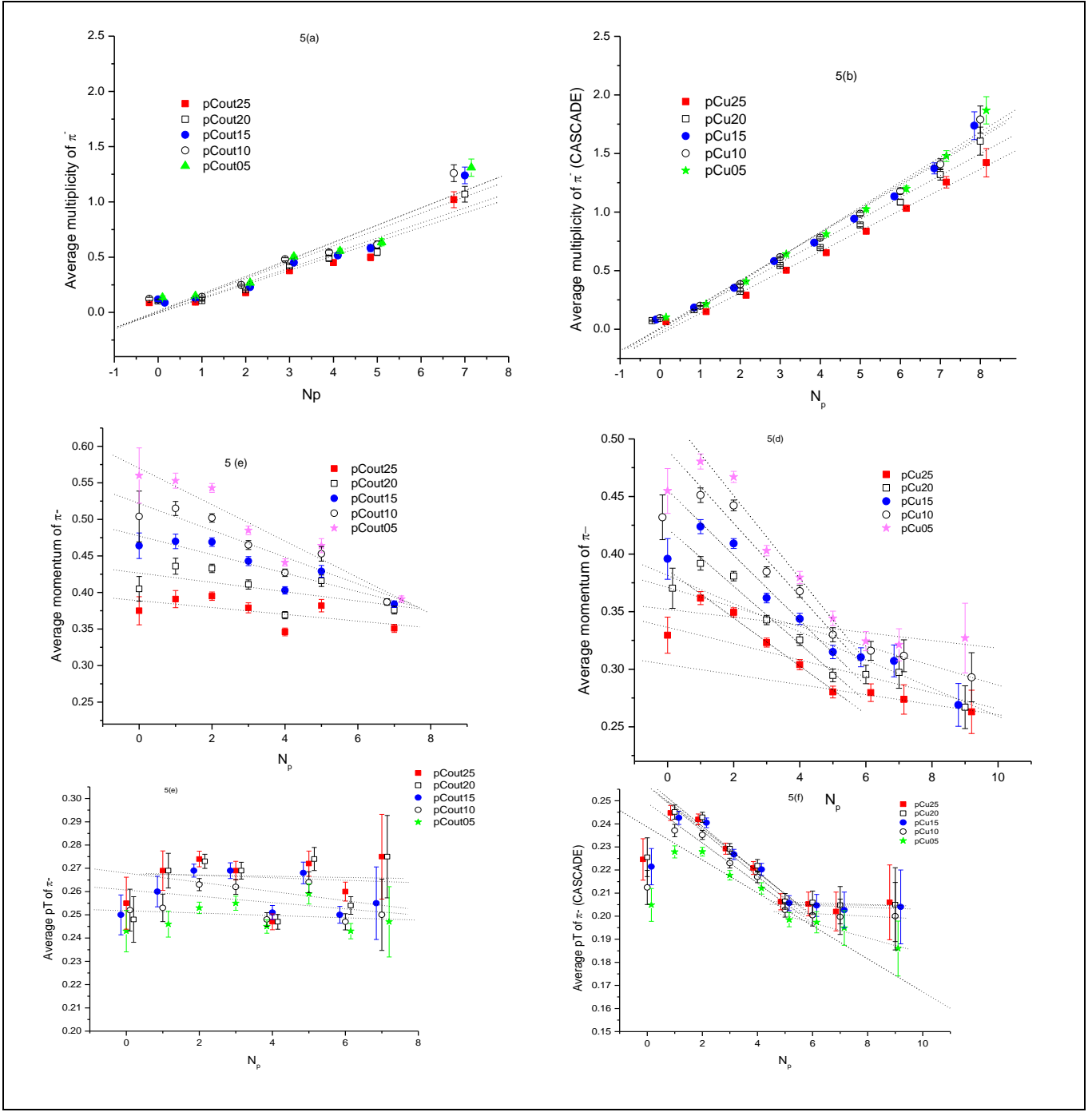

*Figure 5 The average multiplicity, average momentum and average transverse momentum of $\pi^-$ - meson outercone from experimental data in pC collision at 4.2A GeV/c (left hand side from top to bottom) and from Cascade model in pC collision at 4.2A GeV/c (right hand side from top to bottom) as a function of number of identified protons* with $\theta_{½}$=25(red square), $\theta_{½}$=20(open square), $\theta_{½}$=15(Blue triangles), $\theta_{½}$=10(open triangles) and $\theta_{½}$=05(Green stars) *as indicated in the figures.*

So we could say that the experimental data on behavior of the average characteristics of the out cone $\pi^-$-mesons demonstrate some transparency which could not be described by cascade model. This behavior could not be the reason of leading effect due to the fact that the out cone negative pions are secondary particles having large value of angle and small value of momentum.

Table 5. The values of the parameter B for out cone $\pi^-$ -mesons emitted in the pC-interactions at 4.2 A GeV/c.

| $\theta_{1/2}$ | <n> | | <p> | | | <p_T> | | |
|---|---|---|---|---|---|---|---|---|
| | Exp. | cascade | Exp. | Cas. $N_p<5$ | Cas. $N_p>5$ | Exp. | Cas. $N_p<5$ | Cas. $N_p>5$ |
| 5 | 0.16±0.02 | 0.21±0.01 | -0.025±0.003 | -0.036±0.004 | -0.004±0.004 | -0.0005±0.0010 | -0.007±0.001 | -0.0033±0.0006 |
| 10 | 0.16±0.02 | 0.21±0.01 | -0.018±0.003 | -0.032±0.004 | -0.008±0.008 | -0.00135±0.001 | -0.009±0.001 | -0.0006±0.0004 |
| 15 | 0.15±0.02 | 0.20±0.01 | -0.013±0.003 | -0.028±0.003 | -0.012±0.003 | -0.00198±0.002 | -0.009±0.001 | -0.0004±0.0004± |
| 20 | 0.14±0.02 | 0.190±0.009 | -0.006±0.004 | -0.025±0.002 | -0.007±0.003 | -0.000536±0.002 | -0.010±0.001 | -0.0004±0.0002 |
| 25 | 0.13±0.02 | 0.180±0.006 | -0.005±0.003 | -0.021±0.001 | 0.0043±0.0008 | -0.0002±0.0020 | -0.010±0.001 | -0.0001±0.0008 |

The values of outer cone $\pi^-$ -mesons' average multiplicity, average momentum and average transverse momentum from experimental data in dC-interactions at 4.2 A GeV/c as a function of $N_p$ are shown in figure 6 (same notations with previous Figures are used). To compare the results with pC-interactions' data of figure 5 the same characteristics are checked at the following values of $\theta_{1/2} = 5^o$, $10^o$, $15^o$, $20^o$ and $25^o$ as before. The values of the $<n^{out}_{\pi^-}>_{dC}$ at different half angle $\theta_{1/2}$ as a function of the $N_p$ are shown in Fig. 6(a) and 6(b) for experimental and code data respectively. The value $<n^{out}_{\pi^-}>_{dC}$ increase linearly with $N_p$ for both experimental and code data in the dC collision as was the case with the pC collision in figure 5(a) and 5(b). The average multiplicity in case of the Cascade model again has a steeper slope as compared to experimental data. Increasing $\theta_{1/2}$ the slope in the two cases decreases.

The Fig. 6(c) and 6(d) demonstrate the values $<p^{out}_{\pi^-}>_{dC}$ as a function of the $N_p$ for the experimental and code data respectively. There is a big difference between the behavior of the experimental and the code data. Experimental data decrease linearly with $N_p$ and the slope of lines depend on $\theta_{1/2}$. At $\theta_{1/2} = 25^0$ we could see that the slope of line become minimum – transparency, as was the case in pC-interactions. The code data has a steeper slope and is not able to explain the experimental results.

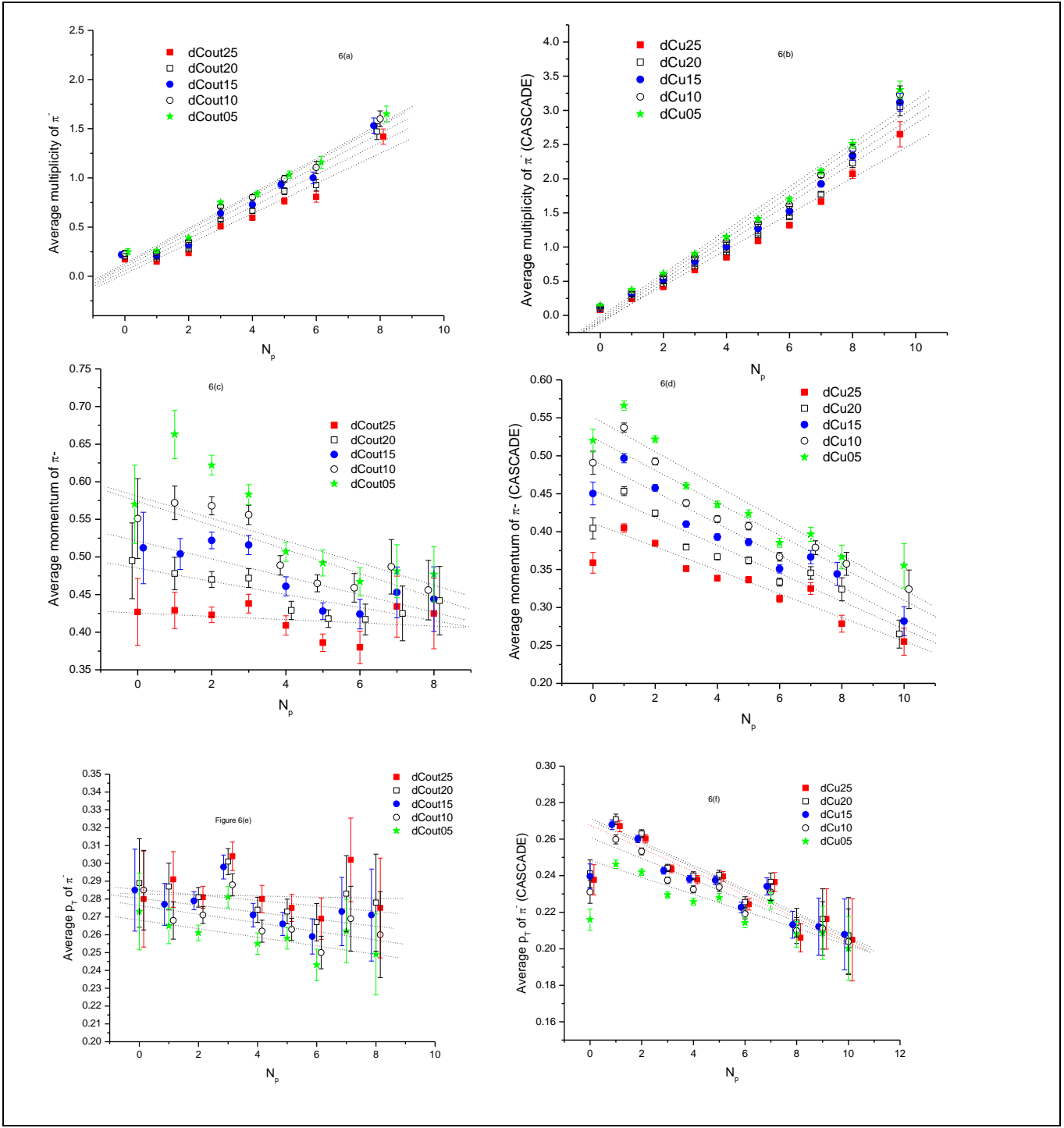

Figure 6 the average multiplicity, average momentum and average transverse momentum of $\pi^-$- meson outercone from experimental data in dC collision at 4.2A GeV/c (left hand side from top to bottom) and from Cascade model in dC collision at 4.2A GeV/c (right hand side from top to bottom) as a function of number of identified protons with $\theta_{1/2}$=25(red square), $\theta_{1/2}$=20(open square), $\theta_{1/2}$=15(Blue triangles), $\theta_{1/2}$=10(open triangles) and $\theta_{1/2}$=05(Green stars) as indicated in the figures.

The Fig. 6(e) and 6(f) demonstrate the average values for $<p_T^{out}{}_{\pi^-}>_{dC}$ as a function of the $N_p$ for the experimental and code data respectively. The last graphs in dC collision again restore the previous results. One can see a huge differences between experimental and code data. Experimental data shows some linear behavior with no dependence on $N_p$. The behavior of $<p_T^{out}{}_{\pi^-}>_{dC}$ is a clear signature of NT as was also the case with $<p_T^{out}{}_{\pi^-}>_{pC}$. The code data shows strong dependence of $<p_T^{out}{}_{\pi^-}>_{dC}$ on $N_p$. Here the $<p_T^{out}{}_{\pi^-}>_{dC}$ value decreases sharply. There is different behavior for different $\theta_{\frac{1}{2}}$ as a function of $N_p$.

Concluding the comparison of the pC- and dC-interactions one can see that the results of dC-interactions data are almost the same as the results of the pC-interactions data (from Figure 5). In both the pC- and dC-interactions the results of the $<n^{out}{}_{\pi^-}>$ have no transparency. But the values of the graphs in experimental data have far less slope than the code data. Furthermore we have observed signal of transparency in the experimental data of $<p^{out}{}_{\pi^-}>$ in both pC and dC at some angles which the code could not explain. Also the transparency of the $<p_T^{out}{}_{\pi^-}>_{dC}$ in experimental data could not be explained by the Cascade model. This type of transparency may be the result of the collective behavior and may be connected to the medium properties. As this kind of transparency could not be explained as a result of leading effect because mesons are newly produced particles and the effect also could not be explained as a result of cascade mechanism because code data could not describe the behaviors of the distributions.

Table 6. The values of the parameter of the B for out cone $\pi^-$ - mesons produced in dC-interactions at 4.2 A GeV/c.

| $\theta_{\frac{1}{2}}$ | <n> | | <p> | | <p_T> | |
|---|---|---|---|---|---|---|
| | experiment | cascade | experiment | cascade | Experiment | cascade |
| 5 | 0.18±0.01 | 0.32±0.02 | -0.015±0.004 | -0.023±0.003 | -0.0024±0.0009 | -0.0047±0.0006 |
| 10 | 0.18±0.01 | 0.31±0.02 | -0.016±0.003 | -0.021±0.003 | -0.002±0.001 | -0.0058±0.0007 |
| 15 | 0.17±0.01 | 0.30±0.02 | -0.012±0.003 | -0.021±0.002 | -0.0020±0.0008 | -0.0064±0.0007 |
| 20 | 0.16±0.01 | 0.29±0.02 | -0.009±0.002 | -0.018±0.002 | -0.0015±0.0008 | -0.0065±0.0007 |
| 25 | 0.15±0.02 | 0.26±0.02 | -0.002±0.003 | -0.016±0.001 | -0.0003±0.0015 | -0.0065±0.0008 |

Before the conclusion we present two tables which give information on appearance of the transparency in our investigations. Table 7 and 8 given below shows all the results on the study of inner and outer cone average multiplicity, average momentum and average transverse momentum of protons, $\pi^-$-meson and $\pi^+$- meson from experimental as well as Cascade model in pC and dC collision at 4.2A GeV/c as a function of number of identified protons with different values of $\theta_{\frac{1}{2}}$. The different symbols used have the following meanings.

+L: explanation in terms of leading effect only

+C: explanation in terms of Cascade model only

+M explanation in terms of Medium effect only

-: negative sign means no such effect observed

+M+C: effect which could be explained in terms of both Medium and cascade effect

$N_p>3$: means effect if observed only with $N_p>3$

Table 7 Results of inner and outer cone average multiplicity, average momentum and average transverse momentum of protons, $\pi^-$ meson and $\pi^+$- meson from experimental as well as Cascade model in pC and dC collision at 4.2A GeV/c as a function of number of identified protons with different values of $\theta_{½}$.

| Inner cone | proton | | | π- | | | π+ | | |
|---|---|---|---|---|---|---|---|---|---|
| Coll. | <n> | <p> | <p_T> | <n> | <p> | <p_T> | <n> | <p> | <p_T> |
| pC | +L | - | +C | +C | +C | +C | +C (Np>3) | +M | +C |
| dC | +L (Np>2) | - | +L+C | +C | +C | +C | - | - | +C |
| Outer cone | proton | | | π- | | | π+ | | |
| Coll. | <n> | <p> | <p_T> | <n> | <p> | <p_T> | <n> | <p> | <p_T> |
| pC | - | +C | - | - | +M | +M | +M+C | +M | +M |
| dC | - | - | - | - | +M | +M | - | - | +M >3 |

## Conclusion: -

1. The behaviors of average multiplicity, 3 momentum in lab frame and transvers momentum for protons and pions were studied as a function of identified protons in pC- and dC-interactions at 4.2 A GeV/c using half angle technique.
2. The identified protons and pions were divided into two groups depending on their angle in lab. frame and particle with angle: less than half angle were considered as inner cone particles; greater than half angle were taken as out cone particles.
3. The values of half angles used were $5^0$; $10^0$; $15^0$; $20^0$; $25^0$ (the angle which divides all secondary particles produced in nucleon-nucleon collisions at 4.2 A GeV/c in two equal parts).

4. The results were approximated using linear function and compared with the data coming from Dubna Cascade Model.
5. We observed several cases for which behaviors of average multiplicity, 3 momentum in lab frame and transvers momentum didn't depend on the number of identified protons – some signals on appearance of nuclear transparency effect.
6. The signals were characterized in three groups of transparency:

I. Transparency due to leading effect: projectile gives some part of its energy during interaction and could save other essential part of its energy. The particle will have maximum energy in an event, which passes very fast by the medium. Such particle cannot interact more and that is why medium seems transparent for it.

II. Cascade Transparency because data coming from the code could satisfactorily describe the effects.

III. Transparency which could not be explained as a result of leading effect and as a result of cascade mechanism.

7. The investigation is going on; we try to understand if the last effect from 6 reflects really some particular properties of the strongly interacting medium.

## References: -